\documentclass[12pt]{article}
\usepackage{a4wide,epsfig}
\usepackage{scalefnt}
\usepackage[latin1]{inputenc}
\usepackage{amsmath}
\usepackage{amsfonts}
\usepackage[english]{babel}

\def\lQ{\Lambda_{\rm QCD}}
\newcommand{\nn}{\nonumber}
\newcommand{\be}{\begin{equation}}
\newcommand{\ee}{\end{equation}}
\newcommand{\bea}{\begin{eqnarray}}
\newcommand{\eea}{\end{eqnarray}}

\def\als{\alpha_{\rm s}}
\def\siml{{\ \lower-1.2pt\vbox{\hbox{\rlap{$<$}\lower6pt\vbox{\hbox{$\sim$}}}}\ }} 

\newcommand{\MS}{\overline{\rm MS}}

\begin{document}

\begin{flushright}
  UAB-FT-630 \\
 UB-ECM-PF-07-07
\end{flushright}
\vspace*{1cm}
\begin{center}
  {\sc \large  Constraints on Regge models from perturbation theory} \\
   \vspace*{2cm} {\bf Jorge~Mondejar$^a$
and Antonio~Pineda$^b$}\\
\vspace{0.6cm}
{\it $^a$\ Dept. d'Estructura i Constituents de la Mat\`eria\\
U. Barcelona, Diagonal 647, E-08028 Barcelona, Spain
\\
[10pt]
$^b$\ Grup de F\'\i sica Te\`orica and IFAE, Universitat
Aut\`onoma de Barcelona, E-08193 Bellaterra, Barcelona, Spain\\}
  \vspace*{2.4cm}
  {\bf Abstract} \\
 \end{center}
We study the constraints that the operator product expansion 
imposes on large $N_c$ inspired QCD models for current-current 
correlators. We focus on the constraints obtained by going 
beyond the leading-order parton computation. 
We explicitly show that, assumed a given mass spectrum: linear Regge 
behavior in $n$ (the principal quantum number) plus corrections 
in $1/n$, we can obtain the logarithmic (and constant) behavior in $n$ 
of the decay constants within a systematic expansion in $1/n$. Our 
example shows that it is possible to have different large $n$ behavior for the 
vector and pseudo-vector mass spectrum and yet comply with all the constraints from the
operator product expansion.  
\\[2mm]

\vspace*{5mm}
\noindent

\newpage


\section{Introduction}

The  operator product expansion (OPE) has been used since long in order to gain information on the non-perturbative 
dynamics of the hadronic spectrum and decays 
\cite{Beane:2001uj,Golterman:2001pj,Cohen:2002st,Golterman:2002mi,Afonin:2004yb,Sanz-Cillero:2005ef,Shifman:2005zn,Cata:2006fu,Afonin:2006da}. 
In this article we revisit this problem. We want to obtain the constraints that the knowledge of the perturbative 
expansion in $\als(Q^2)$ of the current-current correlators in the 
Euclidean regime poses on the relation between the decay constants and the mass spectrum for excitations 
with a large quantum number $n$ (where $n$ is the 
the quantum number of the bound state). We put special emphasis in going 
beyond the leading-order parton computation.
We will work with a specific model for the hadronic spectrum. This is compulsory, since 
different spectral functions\footnote{In particular the one derived directly from perturbation theory, which we do not consider, 
since we will work in the large $N_c$ limit with infinitely narrow resonances.} 
may yield the same OPE expression, yet we believe some aspects of our discussion may 
hold beyond the assumptions of our model.

In order to have a well defined bound state it is crucial to consider the large $N_c$ approximation \cite{hooft1}. 
This ensures infinitely narrow resonances at arbitrarily large energies. We will consider to be 
in the large $N_c$ limit in what follows, as well as in the exact chiral (massless) limit. We will then set 
a specific model for the hadronic spectrum, valid for large values of $n$ (we only need the behavior of the 
spectrum and decays for large $n$, we do not aim to get any information from perturbation theory for low values of $n$). 
This model will be based on the Regge behavior plus corrections in $1/n$ 
that will be included in a systematic way. 
The model is based on the assumption that the Regge behavior is a good description of the spectrum for large $n$ 
(this can be explicitely seen in the 't Hooft model \cite{hooft2} and it is also consistent with phenomenology). 
Given the $1/n$ corrections to the mass spectrum, the expression of the correlator can also be written as  a systematic expansion in $1/n$, where higher 
powers in $1/n$ are equivalent to higher orders in $1/Q^2$ in its OPE. 
By matching the OPE and hadronic expressions order by order in $1/Q^2$, we will be able to predict the logarithmic dependence on $n$
of the decay constants (actually also the constant terms). This result can also be systematically organized within an expansion in $1/n$ together with an 
expansion in $1/\ln n$. We will give explicit expressions up to order $1/n^2$ and $1/\ln^3n$. We will also make some 
numerical estimates of the impact of these corrections. Finally, we would like to stress that we are able to 
 introduce power corrections 
in $1/n$ to the Regge behavior and yet comply with the OPE. This is in contrast with Ref. \cite{Afonin:2004yb}, 
where, besides the Regge behavior, only exponentially suppressed terms are introduced 
(parametrically smaller than any finite power of $1/n$ for large $n$). 
This parameterization is however fine if considered 
as a fit not emanated from the large $n$ limit.

\section{Correlators}

For definiteness, we will consider the vector-vector correlator 
but most of the discussion applies to any other current-current correlator (axial-vector, scalar, ....).
\be
\Pi_V^{\mu\nu}(q)
\equiv
(q^{\mu}q^{\nu}-g^{\mu\nu}q^2)\Pi_V(q)
\equiv
i\int d^4x e^{iqx}\langle vac|T\left\{J_V^{\mu}(x)J_V^{\nu}(0)\right\}|vac\rangle
\,,
\ee
where $J_V^{\mu}=\sum_fQ_f\bar \psi_f\gamma^{\mu}\psi_f$.
In order to avoid divergences, we will consider the Adler function 
\be
{\cal A}(Q^2)\equiv -Q^2{d \over dQ^2}\Pi(Q^2)=
Q^2\int_0^{\infty}dt\frac{1}{(t+Q^2)^2}\frac{1}{\pi}{\rm Im}\Pi_V(t)
\,,
\ee
where $Q^2=-q^2$ is the Euclidean momentum.

Since we are working in the large $N_c$ limit, the spectrum  
consists of infinitely narrow resonances, and the Adler function can be written 
in the following way
\be
{\cal A}(Q^2)=Q^2\sum_{n=0}^{\infty}
{F_V^2(n) \over (Q^2+M_V^2(n))^2}
\,.
\ee 

On the other hand, for large positive $Q^2$, 
one may try to approximate the Adler function by 
its OPE, 
which reads
\bea
{\cal A}_{OPE}(Q^2)
&=&
\sum_fQ_f^2\left[\frac{4}{3}\frac{N_c}{16\pi^2}
\left(
1+\frac{3}{8}N_c\frac{\alpha_{\cal A}(Q^2)}{\pi}
\right)
\right.\label{AOPE}
\\
\nn
&&
\left.
+\frac{C(\als(Q^2))}{Q^4} \beta(\als(\nu))\langle vac |G^2(\nu)| vac \rangle
+{\cal O}\left(\frac{1}{Q^6}\right)
\right]
\,,
\eea
where $\alpha_{\cal A}(Q^2)$ admits an analytic expansion in terms 
of $\als(Q^2)$ (computed in the $\MS$ scheme),
\be
\beta(\als)=-\beta_0\frac{\als(Q^2)}{4\pi}-\beta_1\left(\frac{\als(Q^2)}{4\pi}\right)^2+\cdots
\, ,
\ee
 with $\beta_0=11/3N_c$, $\beta_1=34/3N_c^2$, $\beta_2=2857/54N_c^3$, 
and \cite{Surguladze:1990sp}
\be
C(\als(Q^2))=-\frac{2}{11N_c}\left(1-\frac{35}{22}N_c\frac{\als(Q^2)}{4\pi}+\cdots\right)
\,.
\ee

\section{Matching}

High excitations of the QCD spectrum are believed to satisfy linear Regge 
trajectories:
$$
\displaystyle{\lim_{n \rightarrow \infty}} \frac{M_{V,n}^2}{n}= \; {\rm constant}
.$$ 

For generic current-current correlators, such behavior 
is consistent with perturbation theory in the Euclidean region at leading order in 
$\als$ if the decay constants are taken to be ``constants", ie. independent 
of the principal quantum number $n$. 

The inclusion of subleading effects in $\als$ can be incorporated into this 
model by changing the $n$ dependence of the decay constants without changing 
the ansatz for the spectrum. The inclusion of these effects has consequences 
on subleading sum-rules and the relation with the non-perturbative condensates.

Here we would like to go beyond the analysis at leading order in $\als$, as well 
as to consider power-like corrections in $1/n$.
We will consider that the large $n$ expression for the mass spectrum can 
be organized within a $1/n$ expansion in a systematic way starting from the 
asymptotic linear Regge behavior. In order to fix (and simplify) the problem we will assume 
that no $\ln n$ term appears in the mass spectrum\footnote{This is a simplification. If 
one considers, for instance, the 't Hooft model \cite{hooft2}, $\ln n$ terms do indeed appear.}.  
Therefore, we write the mass spectrum in the following way (for large $n$)
\be
\label{massn}
M_{V}^2(n)=\sum_{s=-1}^{\infty}B_V^{(-s)}n^{(-s)}=B_V^{(1)}n+B_V^{(0)}+\frac{B_V^{(-1)}}{n}+\cdots
\,,
\ee
where $B_V^{(-s)}$ are constants. We will usually denote $M_{V,LO}^2(n)=B_V^{(1)}n$, 
 $M_{V,NLO}^2(n)=B_V^{(1)}n+B_V^{(0)}$ and so on. To shorten the notation, we 
will denote $B_V^{(1)}=B_V$, $B_V^{(0)}=A_V$ and $B_V^{(-1)}=C_V$.

For the decay constants, we will have a double expansion in $1/n$ and $1/\ln n$.  
\be
\label{decayn}
F_{V}^2(n)=\sum_{s=0}^{\infty}F_{V,s}^{2}(n)\frac{1}{n^s}=F_{V,0}^{2}(n)+\frac{F_{V,1}^{2}(n)}{n}+\frac{F_{V,2}^{2}(n)}{n^2}+\cdots
\,,
\ee
where the coefficients $F_{V,s}^{2}(n)$ have a logarithmic dependence on $n$:
\be
F_{V,s}^{2}(n)= \sum_{r=0}^{\infty}C_{V,s}^{(r)}(n)\frac{1}{\ln^r n} \ .
\ee
As we did with the masses, we will define $F_{V,LO}^2(n)=F_{V,0}^{2}(n)$,  
$F_{V,NLO}^2(n)=F_{V,0}^{2}(n)+\frac{F_{V,1}^{2}(n)}{n}$, and so on. Note that 
in this case we also have an expansion in $1/\ln n$. 

We are now in position to start the computation. Our aim is 
to compare the hadronic and OPE expressions of the Adler function 
within an expansion in $1/Q^2$, but keeping the logarithms of $Q$. 
In order to do so we have to arrange the hadronic expression appropiately. Our 
strategy is to split the sum over hadronic resonances into two pieces, for 
$n$ above or below some arbitrary but formally large $n^*$, such that 
$\lQ n^* \ll Q$. The sum up to $n^*$ can be analytically expanded in $1/Q^2$ 
and will not generate $\ln Q^2$ terms (neither a constant term at leading 
order in $1/Q^2$). For the sum from $n^*$ up to $\infty$, we can use Eqs.  (\ref{massn}) and (\ref{decayn})
and the Euler-MacLaurin formula to transform the sum in an integral plus 
corrections in $1/Q^2$. Whereas the latter do not produce logarithms, the integral does. 
These logarithms of $Q$ are generated by the large $n$ behavior of the bound 
states and the introduction of powers of $1/n$ is equivalent (once 
introduced in the integral representation, and for large $n$) to the 
introduction of (logarithmically modulated) $1/Q^2$ corrections in the OPE expression. 

Therefore, by using the Euler-MacLaurin formula, we write the Adler function in the
following way ($B_2=1/6$, $B_4=-1/30$, ...)
\bea
\nn
{\cal A}(Q^2)&=&Q^2\int_0^{\infty} dn {F_V^2(n) \over (Q^2+M_V^2(n))^2}+
Q^2\left[\sum_{n=0}^{n^*-1}{F_V^2(n) \over (Q^2+M_V^2(n))^2}
- \int_0^{n^*}{F_V^2(n) \over (Q^2+M_V^2(n))^2}\right]
\\
&&
\left.
+
{Q^2 \over 2}{F_V^2(n^*) \over (Q^2+M_V^2(n^*))^2}
+
Q^2\sum_{k=1}^{\infty}(-1)^k{|B_{2k}| \over (2k)!}{d^{(2k-1)} \over dn^{(2k-1)}}
{F_V^2(n) \over (Q^2+M_V^2(n))^2}\right|_{n=n^*}
\,,
\label{Ahadr}
\eea
where $n^*$ stands for the subtraction point we mentioned above, such that 
for $n$ larger than $n^*$ one can use the asymptotic expressions (\ref{massn}) and 
(\ref{decayn}). This allows us to eliminate terms that vanish when 
$n \rightarrow \infty$. Note that the last sum in Eq. (\ref{Ahadr}) is an asymptotic series, and 
in this sense the equality should be understood. 

Note also that for $n$ below $n^*$, we will not distinguish between LO, NLO, etc...
in masses or decay constants, since for those states we will not assume that one can do an expansion 
in $1/n$ and use Eqs. (\ref{massn}) and 
(\ref{decayn}).

Finally, note that the expressions we have for the masses and decay constants become more and 
more infrared singular as we go to higher and higher orders in the $1/n$ expansion. 
This is not a problem, since we always cut off the integral for $n$ smaller 
than $n^*$. Either way, the major problems would come from the decay constants, since, in the case of 
the mass, $Q^2$ effectively acts as an infrared regulator.

\subsection{LO Matching}

We want to match the hadronic, Eq. (\ref{Ahadr}), and OPE, Eq. (\ref{AOPE}), expressions for the Adler function at the 
lowest order in $1/Q^2$. This means that we have to consider the lowest order expressions in $1/n$ for the 
masses and decay constants, i.e. $F_{V,LO}^2(n)$ and $M_{V,LO}^2(n)$,  
since the corrections in $1/n$ give contributions suppressed by powers of $1/Q^2$.

Only the first term in Eq. (\ref{Ahadr}) can generate logarithms or terms that 
are not suppressed by powers of $1/Q^2$. Therefore we obtain the following equality,
\be
{\cal A}^{pt.}(Q^2) 
\equiv Q^2\int_0^{\infty}dn \frac{F_{V,LO}^2(n)}{(Q^2+M_{V,LO}^{2})^2}
=
\frac{4}{3}\frac{N_c}{16\pi^2}\sum_fQ_f^2
\left(
1+\frac{3}{8}N_c\frac{\alpha_{\cal A}(Q^2)}{\pi}
\right)
\,.
\ee
This equation can be fulfilled by demanding that
\be
{F_{V,LO}^2(n) \over |d M^2_{V,LO}(n)/dn|}
=
{1 \over \pi}{\rm Im} \Pi^{pert.}_{V}(M^2_{V,LO}(n))
\,.
\ee
By using the perturbative expression for ${\rm Im} \Pi^{pert.}_{V}$ (see \cite{Chetyrkin:1996ez}), we obtain
\begin{eqnarray}
F_{V,LO}^2(n)&=&B_V\frac{4}{3}\frac{N_c}{16\pi^2}\sum_fQ_f^2\left\{1+\frac{3}{8\pi}\,
N_c\als(nB_V)+\frac{243-176\,\zeta(3)}{128\pi^2}\,N_c^2\als^2(nB_V) \right.\\
&&\left. +\frac{346201-2904\pi^2-324528\,\zeta(3)+63360\,\zeta(5)}{27648\pi^3}\,
N_c^3\als^3(nB_V)
+
{\cal O} \left(\als^4(nB_V) \right) \right\} \ , \nonumber
\end{eqnarray}
where $\als(nB_V)$ should actually be understood as a function of $\als(B_V)$ 
and $\ln n$. Therefore, 
it is obvious that the above expression is resumming powers of $\als(B_V)\ln n$: 
\begin{eqnarray}
F_{V,LO}^2(n)&=&B_V\frac{4}{3}\frac{N_c}{16\pi^2}\sum_fQ_f^2
\left\{1+\frac{3}{2}\frac{1}{1+\frac{11}{3}N_c\frac{\als(B_V)}{4\pi}\ln(n)}\,
N_c\frac{\als(B_V)}{4\pi}\right. \\
&&+\frac{\left(2673-1936\,\zeta(3)-408\,
\ln\left(1+\frac{11}{12\pi}N_c\als(B_V)\ln(n)\right)\right)}
{88(1+\frac{11}{3}N_c\frac{\als(B_V)}{4\pi}\ln(n))^2}
\,\frac{N_c^2\als^2(B_V)}{(4\pi)^2} \nonumber \\
&& +\frac{N_c^3\als^3(B_V)}{(4\pi)^3}\frac{1}{52272\pi(1+\frac{11}{3}N_c\frac{\als(B_V)}{4\pi}
\ln(n))^3}
\left[-350427\,N_c\als(B_V)\ln(n)
\right.
\nn
\\
&&
+121\pi\left(346201-2904\pi^2
\nonumber
-324528\,\zeta(3)+63360\,\zeta(5)\right)
\\
&&
-3672\pi(2877-1936\,\zeta(3))\ln\left(1+\frac{11}{12\pi}\als(B_V)\ln(n)\right) \nn \\
&& 
\left.\left.
+749088\pi\,\ln^2\left(1+\frac{11}{12\pi}\als(B_V)\ln(n)\right)\right]\nn 
+
{\cal O} \left(\als^4(B_V) \right)\right\} \ . 
\end{eqnarray}
Doing so we see that we are able to obtain the dependence of the 
decay constant in $\ln n$ (somewhat we are assuming that $\als(B_V)$ is an 
small parameter, $B_V \sim 1$ GeV).

We can also rewrite the decay constant as an expansion in $1/\ln n$ by using the 
equality
\begin{equation}
\label{alpha}
\ln \tilde n 
=
\frac{1}{\beta_0}\left(\frac{4\pi}{\als(nB_V)}
+
\frac{\beta_1}{\beta_0}ln\left(\beta_0\, \frac{\als(nB_V)}{4\pi}\right) 
+ \left(\frac{\beta_2}{\beta_0}-\left(\frac{\beta_1}{\beta_0}\right)^2\right)\frac{\als(nB_V)}{4\pi}\right) \,
\,,
\end{equation}
where ${\tilde n}=nB_V/\Lambda_{\MS}$. We then obtain
\bea
\nn
F_{V,LO}^2(n) &=&B_V{4 \over 3}{N_c \over 16 \pi^2}\sum_fQ_f^2
\left\{
1+{9 \over 22}{1 \over \ln {\tilde n}}
+{1 \over \ln^2 {\tilde n}}
\left[
-{459 \over 1331} \ln \ln {\tilde n}
+{144 \over 121}
\left({243 \over 128}-{11 \over 8}\zeta(3)\right)
\right]
\right.
\\
\nn
&&
+{1 \over \ln^3 {\tilde n}}
\left[
{46818 \over 161051}\ln^2 \ln {\tilde n}
+{459 \over 322102}(-2877+1936\zeta(3))\ln \ln {\tilde n}
+{42272605 \over 2576816}
\right.
\\
\label{pert}
&&
\left.
\left.
-\frac{3\,{\pi }^2}{22} - 
  \frac{20283\,{\zeta}(3)}{1331} + 
  \frac{360\,{\zeta}(5)}{121}
\right]
+{\cal O}\left(\frac{1}{\ln^4 n}\right)
\right\}
\,.
\eea

Note that the lowest contribution in $1/\ln n$ for the decay constant,
$B_V{4 \over 3}{N_c \over 16 \pi^2}\sum_fQ_f^2$, 
 which, usually, is the only one considered,  
reproduces the leading-order partonic prediction for the Adler function. 

Note also that there is no problem with the Landau pole, even if 
the result is written in the form of Eq. (\ref{pert}), since it holds only for $n$ larger than an $n^*$ such that $\Lambda_{\MS} \ll n^*B_V$ (the integral has an infrared cutoff at $n^*$). 

Finally, we remind that, strictly speaking, we can only fix the ratio between the 
decay constant and the derivative of the mass. We have fixed 
this ambiguity by arbitrarily imposing the $n$ dependence of the 
mass spectrum. 

\subsection{NLO matching}

We now want to obtain extra information on the decay constant by demanding the 
validity of the OPE at ${\cal O}(1/Q^2)$, in particular the absence 
of condensates at this order. We then have to use the NLO expressions
for $M_V^2(n)$ and $F_V^2(n)$. With the ansatz we are using for the 
mass at NLO, it is compulsory to introduce (logarithmically modulated) 
$1/n$ corrections to the decay constant if we want this constraint to hold. Note that it is possible 
to shift all the perturbative corrections to the decay constant.

Imposing that the $1/Q^2$ condensate vanishes produces the following sum rule:
\bea
\label{matchingNLO}
&&
A{d \over dQ^{2}}{\cal A}^{pt.}-{A \over Q^2}{\cal A}^{pt.}
+{1 \over Q^2}\left[\sum_{n=0}^{n^*-1}F_V^2(n) 
- \int_0^{n^*}dnF_{V,LO}^2(n) \right]
+
{F_V^2(n^*) \over 2Q^2}
\\
\nn
&&
\left.
+
{1 \over Q^2}
\sum_{k=1}^{\infty}(-1)^k{|B_{2k}| \over (2k)!}{d^{(2k-1)} \over dn^{(2k-1)}}
F_V^2(n) \right|_{n=n^*}
-Q^2 \int_0^{n^*}dn{ F_{V,1}^2(n)/n \over (Q^2+M_{V,LO}^2(n))^2}
\\
\nn
&&
+Q^2 \int_0^{\infty}dn{F_{V,1}^2(n)/n \over (Q^2+M_{V,LO}^2(n))^2}=0
\,.
\eea
This equality should hold independently of the value of $n^*$, which 
formally should be taken large enough so that $\als(n^*B_V) \ll 1$, 
i.e. the limit $\Lambda_{\MS} \ll n^*B_V \ll Q^2$.
Again, the meaning of the asymptotic series appearing in Eq. (\ref{matchingNLO}) 
should be taken with care. If we forget about this potential problem, only a few terms in Eq. 
  (\ref{matchingNLO}) can generate $\ln Q^2$ terms, which should cancel at any order. 
Those are the first two and the last two terms. Actually, the next to last term does 
not generate logarithms, but it allows to regulate possible infrared divergences appearing 
in the calculation.  
Therefore, asking for the cancellation of the $1/Q^2$ suppressed logarithmic terms 
produced by the first two and the last term in Eq. (\ref{matchingNLO}) 
fixes  $F^2_{V,1}$. The non-logarithmic terms should also 
be cancelled but they cannot be fixed from perturbation theory. 

One can actually find an explicit solution to the above constraint for $F^2_{V,1}$ by performing 
some integration by parts. We obtain
\bea
\frac{F_{V,1}^2(n)}{n}&=&\frac{A_V}{B_V}\frac{d}{dn}F_{V,0}^2(n)
\\
\nn
&=&A_V \frac{4}{3}\frac{N_c}{16\pi^2}\sum_fQ_f^2\frac{1}{n}\left\{-\frac{9}{22}\frac{1}{\ln^2 \tilde n}
-\left[\frac{459}{1331}\left(1-2\,  \ln\left(\ln \tilde n \right)\right)+\frac{2187}{484}-\frac{36\,\zeta (3)}{11}\right]
\frac{1}{\ln^3 \tilde n} \right. \nonumber \\
& & +\frac{3}{2576816}\left[-45794053+351384\pi^2+41637552\, \zeta (3)-7666560\, \zeta (5) \right. \nonumber \\
& & \left.\left. -3672\, \ln\left(\ln\tilde n\right)\left(-3013+1936\, \zeta (3) +204\, \ln\left(\ln \tilde n \right)\right)\right]
\frac{1}{\ln^4 \tilde n}
+
{\cal O} \left(\frac{1}{\ln^5 \tilde n}\right)
\right\} \ , \nonumber
\eea
or in terms of $\als(nB_V)$ or $\als(B_V)$,
\begin{eqnarray}
\frac{F_{V,1}^2(n)}{n}&=& A _V\frac{4}{3}\frac{N_c}{16\pi^2}
\sum_fQ_f^2\frac{1}{n}\left\{-\frac{11}{32\pi^2}\,N_c^2\als^2(nB_V)
-\frac{2877-1936\,\zeta (3)}{768\pi^3}\,N_c^3\als^3(nB_V) \right.\\
& &\left. 
- \frac{11(376357-2904\pi^2-344112\,\zeta (3)+63360\,\zeta (5) )}{110592\pi^4}\,
N_c^4\als^4(nB_V)
+
{\cal O}\left(\als^5(nB_V)\right)\right\} \ , \nonumber
\end{eqnarray}

\begin{eqnarray}
\frac{F_{V,1}^2(n)}{n}
&=& 
A_V\frac{4}{3}\frac{N_c}{16\pi^2}\sum_fQ_f^2\frac{1}{n}
\left\{-\frac{11}{2}\frac{1}{(1+\frac{11}{3}N_c\frac{\als(B_V)}{4\pi}\ln(n))^2}\,
N_c^2\frac{\als^2(B_V)}{(4\pi)^2} \right.\\
&& -\frac{\left(2877-1936\,\zeta (3)-408\,
\ln\left(1+\frac{11}{12\pi}\,N_c\als(B_V)\ln(n)\right)\right)}
{12(1+\frac{11}{3}\,N_c\frac{\als(B_V)}{4\pi}\ln(n))^3}\,
N_c^3\frac{\als^3(B_V)}{(4\pi)^3} \nonumber \\
&&-\frac{1}{4752\pi(1+\frac{11}{3}\,N_c\frac{\als(B_V)}{4\pi}\ln(n))^4}
N_c^4\frac{\als^4(B_V)}{(4\pi)^4}
\left[-233618\,N_c\als(B_V)\ln(n)
\right.
\nn
\\
&&
\nn
+121\pi\left(376357-2904\pi^2 -344112\,\zeta(3)+63360\,\zeta(5)\right)
\\
&&
\nn
-3672\pi  (3013-1936\,\zeta(3))\, \ln\left(1+\frac{11}{12\pi}\als(B_V)\ln(n)\right)  \nn\\
&&
\left.\left.
+749088\pi\, \ln^2\left(1+\frac{11}{12\pi}\als(B_V)\ln(n)\right)\right]+
{\cal O}\left(\als^5(B_V)\right)
\right\}  \ . \nonumber
\end{eqnarray}

Note that besides the $1/n$ suppression, we also have an extra $\als^2(nB_V)$ suppression.

In principle one could think of the existence of $1/n \times constant$ terms in the decay constant, 
i.e. without any associated logarithm. However, such 
terms produce $\ln(Q^2)$ contributions in the Euclidean regime that do 
not appear in the perturbative computation, so they can be ruled out. 
This appears to be a general statement since  $1/n^m \times constant$ for any $m$ 
integer also produces logarithms. Note that in order to give meaning to these 
integrals it is implicit that the integral over $n$ has an infrared cutoff at $n^*$. 
Nevertheless, the logarithm does not appear to multiply powers of the infrared 
cutoff (as expected).

Finally, we would like to mention that, besides the constraints coming from the 
logarithmic related behavior of the OPE, there is also the constraint from its constant terms, 
which should sum up to zero. Nevertheless, 
for this constraint we cannot give a closed expression. This is due to the fact that the $\ln Q^2$ independent 
terms may receive contributions 
from any subleading order in the $1/n$ expansion of the masses and decay constants. The reason is
 that the decay constant at a given order in $1/n$ is obtained after performing some integration by 
parts, which generate new ($\ln Q^2$-independent) terms that can be $Q^2$ enhanced.
This statement is general and also applies to any subleading power in the $1/Q^2$ matching 
computation.

\subsection{NNLO matching}

We now consider expressions for the mass and decay constants at NNLO.
For the first time we have to consider condensates. Simplifying terms that do not produce logs, 
we obtain the following equation,
\bea
&&
\frac{35}{121}\frac{\als(Q^2)}{4\pi}
\frac{\beta(\als(\nu))\langle vac |G^2(\nu)| vac \rangle}{Q^4}
\\
\nn
&&
\doteq
Q^2
\int_{n^*}^{\infty}\frac{dn}{(Q^2+B_Vn)^2}
\left[
\frac{F_{V,2}^2(n)}{n^2}-\frac{1}{B_V}\frac{d}{dn}\left(\frac{C_VF_{V,0}^2(n)}{n}
+\frac{A_VF_{V,1}^2(n)}{2n}\right)
\right] \ ,
\eea
where $\doteq$ stands for the fact that we can only predict the $\ln Q^2$ dependence. Constant
terms are not fixed by this relation. 

In order to get a more closed expression is convenient to 
use the following equality,
\be
\frac{1}{Q^4}\als(Q^2)
\doteq
Q^2\int_{n^*}^{\infty}\frac{dn}{(Q^2+B_Vn)^2}
\frac{1}{B_Vn^2}\frac{\beta_0}{8\pi}\als^2(nB_V) \ ,
\ee
valid up to terms that do not produce logarithms or those that are subleading.

We get then 
\bea
&&F_{V,2}^2(n)
=
-C_V\frac{4}{3}\frac{N_c}{16\pi^2}
\sum_fQ_f^2
\left\{1+\frac{3}{8\pi}\,
N_c\als(nB_V)
\right.
\nn
\\
&&
+
\left[
\frac{287-176\,\zeta(3)}{128\pi^2}
-\frac{11A_V^2}{64\pi^2B_VC_V}
-\frac{35}{88}
\frac{\beta(\als(\nu))\langle vac |G^2(\nu)| vac \rangle}{B_VC_VN_c^2}
\right]
\,N_c^2\als^2(nB_V) \\
&&\left.
\qquad
+
{\cal O} \left(\als^3(nB_V) \right) \right\} \ . \nonumber
\eea
Note that in this case we only consider up to ${\cal O}(\als^2(nB_V))$ 
corrections, since higher order loops are unknown. The accuracy is set 
by our knowledge of the matching coefficient of the gluon condensate.
Note also that $F_{V,2}^2(n)$ does not have $\als$ suppression. Therefore, for 
low $n$, this contribution could be practically of the same size than, 
formally, more important terms.

\section{Axial versus vector correlators}

The above discussion has been performed for the vector-vector correlator Adler function. It goes without saying that 
we could perform a similar analysis with axial-vector currents, since the perturbative expansions 
for both correlators are equal.  Here it comes an important 
observation. We could change the coefficients for the mass spectrum $B_A \not= B_V$,  
$A_A \not= A_V$, $\cdots$, 
yet we would obtain the same expression for the OPE (at the order we are working, the 
first chiral breaking related effects are ${\cal O}(1/Q^6)$). Therefore, we conclude that the OPE 
does not fix $B_A=B_V$ as it is sometimes claimed in the literature 
\cite{Beane:2001uj,Cohen:2002st}\footnote{Another issue, on which we do not enter, 
is whether some other kind of arguments (relying on the specific model 
used), like semiclassical arguments, may fix those parameters to be equal.}.
Our computation gives 
a specific counter example. Moreover, it is nice to see what the role played by $B_A$ and $B_V$ is
in our case. 
When one goes to the Euclidean regime,  $B_A$ and $B_V$ become renormalization factorization 
scales and, obviously, the physical result does not depend on them (for large $Q^2$ in the Euclidean). 
On the other hand, it is evident that having different constants: $B_A$, $B_V$, \dots produces different physical predictions 
for the masses and decay widhts for vector or axial-vector channels. The point to be emphasized is that 
 $B_A = B_V$ cannot come from an OPE analysis alone.  This point has already been stressed 
in Refs. \cite{Golterman:2002mi,Cata:2006fu}, what we think is novel in our analysis is that we have seen that the inclusion of corrections in 
$\als$ does not affect that conclusion, and that $B_A$ and $B_V$ play 
the role of the renormalization scale in the analogous perturbative analysis in the Euclidean 
regime, and are therefore unobservable.
Finally, we cannot avoid mentioning the analysis of Ref. \cite{Casero:2007ae} where, using AdS/CFT, 
they explicitly find Regge behavior with different slopes for vector and axial-vector channels.  

In any case, even though the constants that characterize the spectrum can be different for 
the vector and axial-vector channel, they have to yield the same expressions for the OPE when 
combined with the decay constants. This produces some relations that we list in what follows. 
We first define $t\equiv B_Vn=B_An'$ and take $n$ and $n'$ as continuous variables. We then obtain 
the following equalities
\be
\frac{F_{V,LO}^2(n)}{B_V}=\frac{F_{A,LO}^2(n')}{B_A}=
{1 \over \pi}{\rm Im} \Pi^{pert.}_{V}(t)\equiv f_0(t)
\,,
\ee
\be
\frac{1}{A_VB_V}\frac{F_{V,1}^2(n)}{n}
=
\frac{1}{A_AB_A}\frac{F_{A,1}^2(n')}{n'}
=
\frac{d}{dt}f_0(t)\,,
\ee
\bea
&&
\frac{1}{B_V}\left[
\frac{F_{V,2}^2(n)}{n^2}-\frac{1}{B_V}\frac{d}{dn}\left(\frac{C_VF_{V,0}^2(n)}{n}
+\frac{A_VF_{V,1}^2(n)}{2n}\right)
\right]
\nn
\\
&&
=
\frac{1}{B_A}\left[
\frac{F_{A,2}^2(n')}{n^{\prime 2}}-\frac{1}{B_A}\frac{d}{dn'}\left(\frac{C_AF_{A,0}^2(n')}{n'}
+\frac{A_AF_{A,1}^2(n')}{2n'}\right)
\right]
\nn
\\
&&=
\frac{1}{t^2}\beta(\als(\nu))\langle vac |G^2(\nu)| vac \rangle
f_1(t) 
\,,
\eea
where
\be
f_1(t)=\frac{35}{121}\frac{\beta_0}{2}\frac{\als^2(t)}{(4\pi)^2}
+\cdots
.
\ee

\section{Numerical Analysis}

\begin{table}[h]
\addtolength{\arraycolsep}{0.1cm}
$$
\begin{array}{|l||c|c|c|c|}
\hline
 & n=1 & n=2 & n=3 & n=4
\\ \hline\hline
M_{\rho} ({\rm I}) & 781.3 (775.5\pm 0.4) & 1440.2 (1459\pm 11) & 1891.8
(1870\pm 20)  & 2257 (2265\pm 40)
\\ \hline\hline
M_{\rho} ({\rm II}) & 771.5 (775.5\pm 0.4) & 1471.7 (1459\pm 11) & 1855
(1870\pm 20)  & 2154.8 (2149\pm 17)
\\ \hline\hline
M_{a_1} & 1235.6 (1230\pm 40)  & 1621.7 (1647\pm 22)  & 1962
(1930^{+30}_{-70}) & 2257.8 (2270^{+55}_{-40})
\\ \hline\hline
F_{V} ({\rm I}) & 156 (156 \pm 1) &  155 & 154 & 153
\\ \hline\hline
F_{V} ({\rm II}) & 185 (156 \pm 1) &  147 & 139 & 135
\\ \hline\hline
F_{A} & 123 (122 \pm 24)  & 137 & 139 & 139
\\  \hline
\end{array}
$$
\caption{{\it We give the experimental values of the masses (in MeV)
and electromagnetic decay
constants (when available) for vector and axial vector particles
(within parenthesis), compared with
the values obtained from the fit. For the vector states we consider two 
possible Regge trajectories that we label {\rm I} and {\rm II} respectively. 
We take $\als(1\, {\rm GeV})=0.5$
and $\beta\langle G^2 \rangle=-(352 \,{\rm MeV})^4$.}}
\label{table}
\end{table}

\begin{figure}
\centering
\includegraphics[width=0.41\columnwidth]{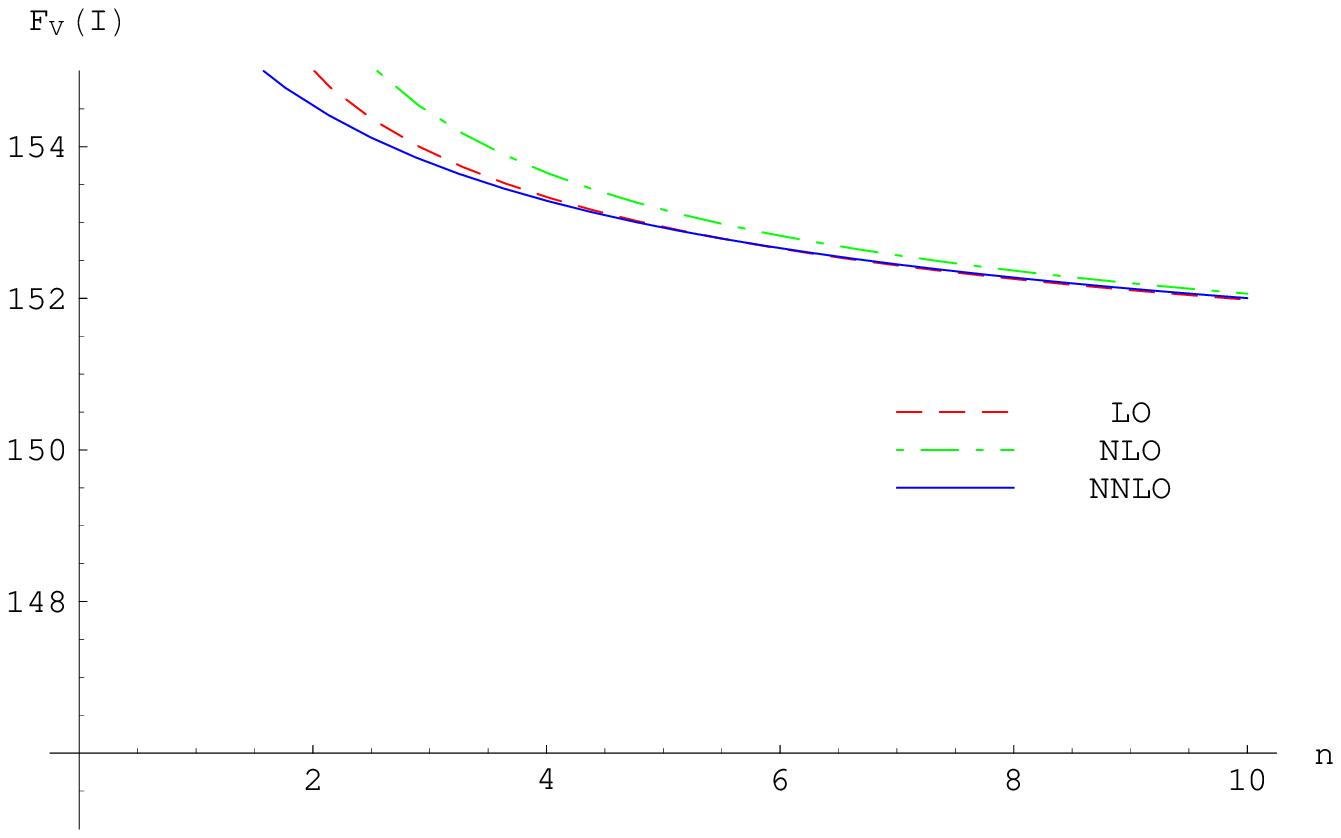}
\hspace{1in}
~\includegraphics[width=0.41\columnwidth]{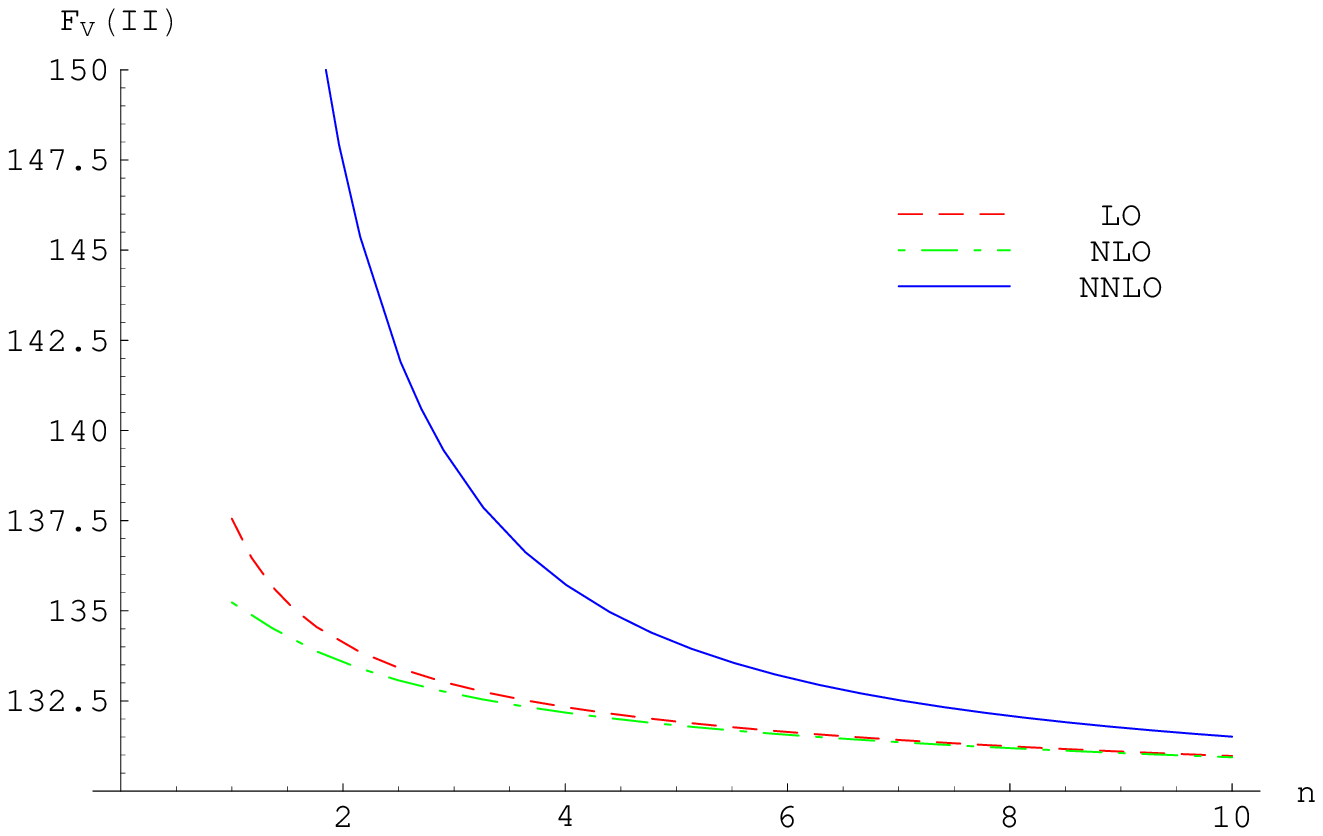}
\caption{\it In this plot we show $F_{V}${\rm(I)} and $F_{V}${\rm(II)} at different orders in the $1/n$ expansion.}
\label{1/n}
\end{figure}

\begin{figure}
\centering
\includegraphics[width=0.41\columnwidth]{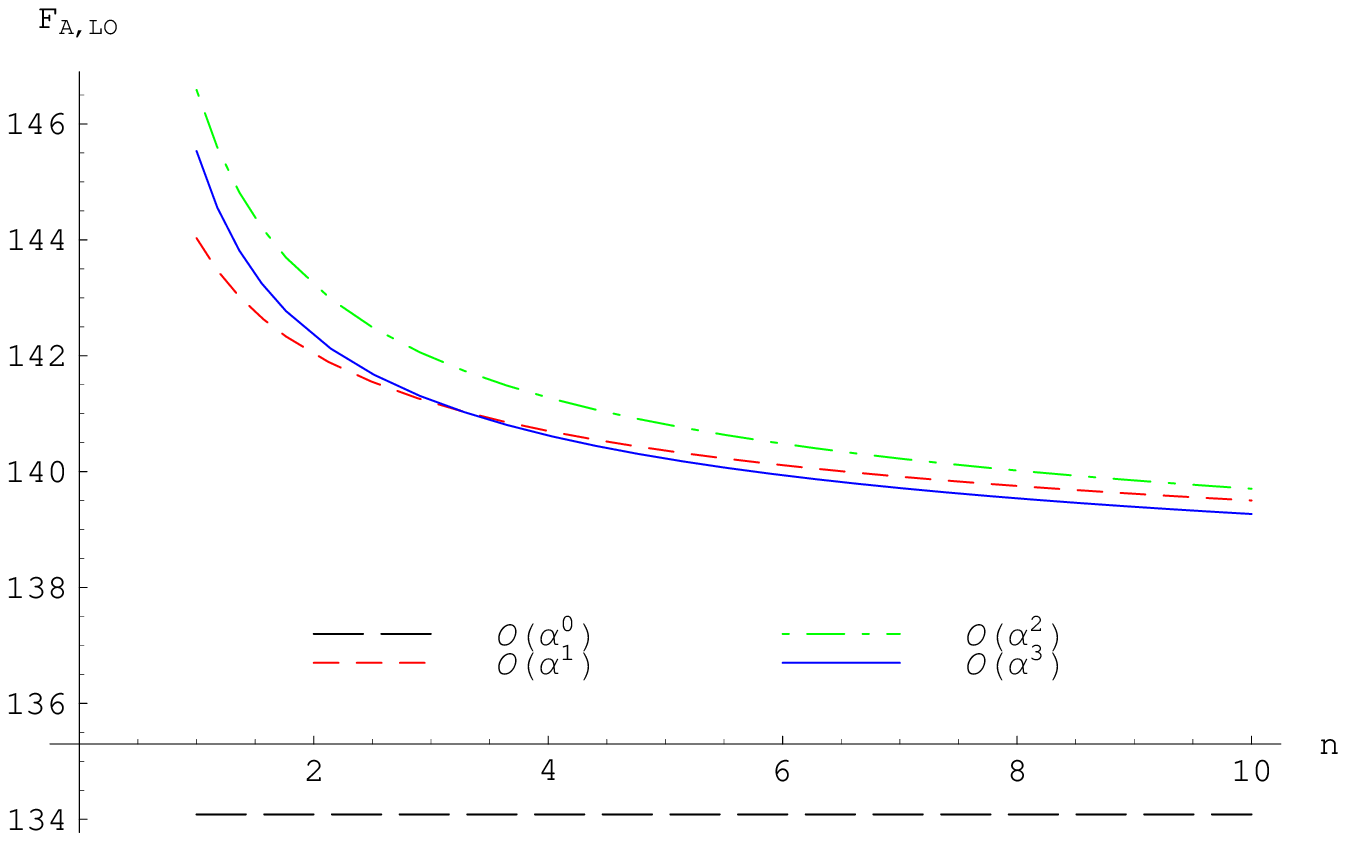}
\hspace{1in}
~\includegraphics[width=0.41\columnwidth]{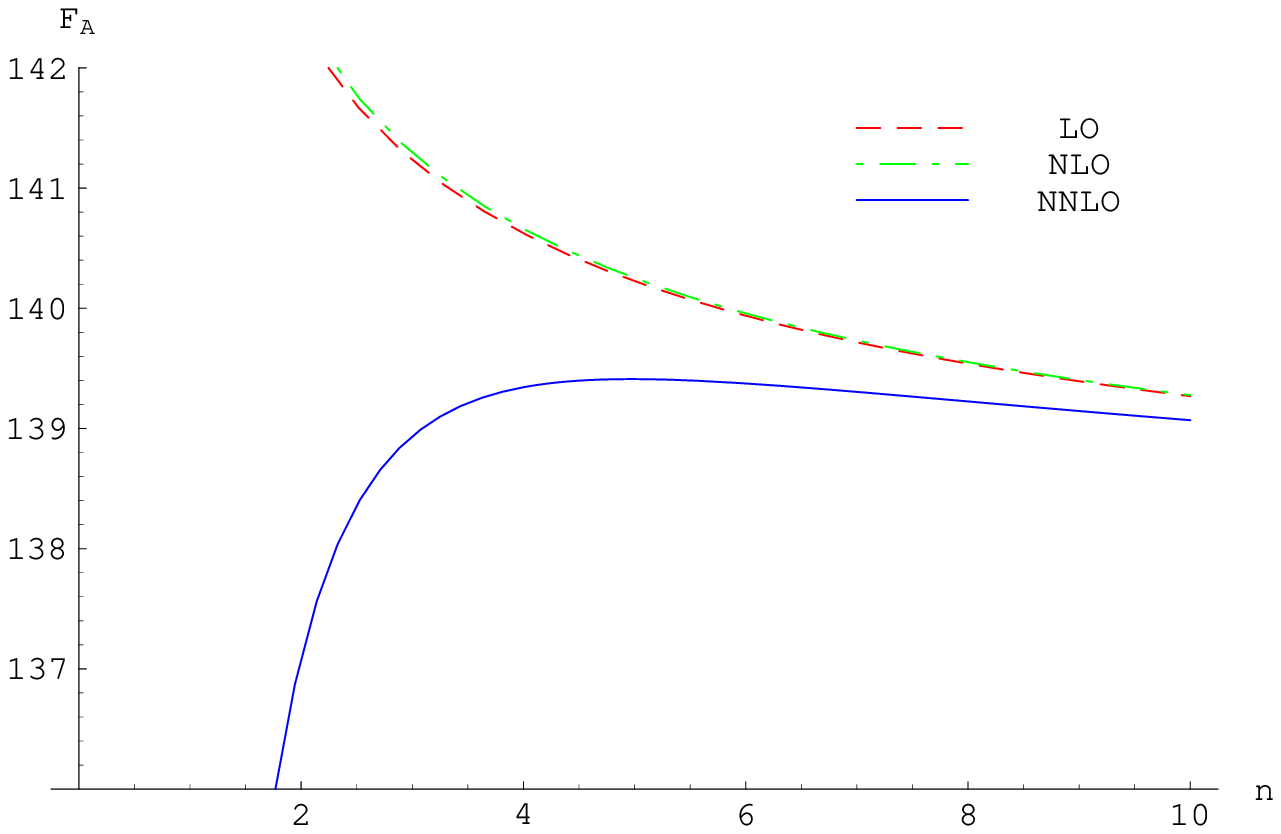}
\caption{\it In this plot we show $F_{A,LO}$ and $F_{A}$  at different orders in $\als$ and in the $1/n$ expansion, respectively.}
\label{A}
\end{figure}

We restrict ourselves to the SU(2) case (non-strange sector) and study the vector and axial-vector channels.
We would like here to assess the importance of including perturbative corrections to a standard 
analysis based on the OPE. We do not aim to perform a full fledged analysis, but only to see the importance 
of the corrections. 
In table \ref{table}, we give the values of the masses and decay constants. 
In Figure \ref{alpha} we show the changes in both $F_{V,LO}$(I) and $F_{V,LO}$(II) as we include 
higher orders in the expansion in $\als$, and in Figure \ref{1/n} the changes in the full $F_V$(I) 
and $F_V$(II) as we include higher orders in  $1/n$. In figure \ref{A} we show the same plots 
for the axial-vector case. We take the experimental values 
from Ref. \cite{Yao:2006px}. In principle there are more states in the particle data book, 
in particular in the vector channel. 
Nevertheless, it is not clear whether they belong to the same Regge trajectory or whether 
they belong to some daughter one, see, for instance, the discussion in Ref. \cite{Afonin:2004yb}. For the 
time being we will disregard the study of other possible (vector) Regge trajectories 
and restrict the analysis to a single trajectory. We will consider the two possibilities listed in 
Table \ref{table}. Our choice of states for the set (I) is motivated by the discussion of 
Ref. \cite{Glozman:2003bt} on the possible formation of multiplets in the case of chiral symmetry restoration. 
The set (II) is based on the assignment of states made in Ref. \cite{Afonin:2004yb} 
(based on the existence of $S$ and $D$-wave daughter trajectories) and in particular on the analysis of 
Ref. \cite{Bugg:2004xu}, where the state 2265 is argued to belong to the $D$-wave Regge trajectory\footnote{We also 
thank S. Afonin for discussions on this point.}.
 
In order to fix the parameters of the mass spectrum we use the experimental values of the masses we list 
in the table. We obtain the values:
\bea
B_V{\rm (I)}&=&1.525\times 10^6 \, {\rm MeV^2}\,,
\quad
A_V{\rm (I)}= -1.038\times 10^6 \, {\rm MeV^2}\,, 
\quad  
C_V{\rm (I)}=0.123 \times 10^6 \,
{\rm MeV^2}\,,
\nn
\\
B_V{\rm (II)}&=&1.128\times 10^6 \, {\rm MeV^2}\,,
\quad
A_V{\rm (II)}= 0.353\times 10^6 \, {\rm MeV^2}\,, 
\quad  
C_V{\rm (II)}= -0.885 \times 10^6 \,
{\rm MeV^2}\,,
\nn
\\
B_A&=&1.278\times 10^6 \, {\rm MeV^2}\,,
\qquad
A_A= -0.100\times 10^6 \, {\rm MeV^2}\,, \qquad  C_A=0.349 \times 10^6 \,
{\rm MeV^2}
\nn
\,.
\\
\eea

\begin{figure}
\centering
\includegraphics[width=0.41\columnwidth]{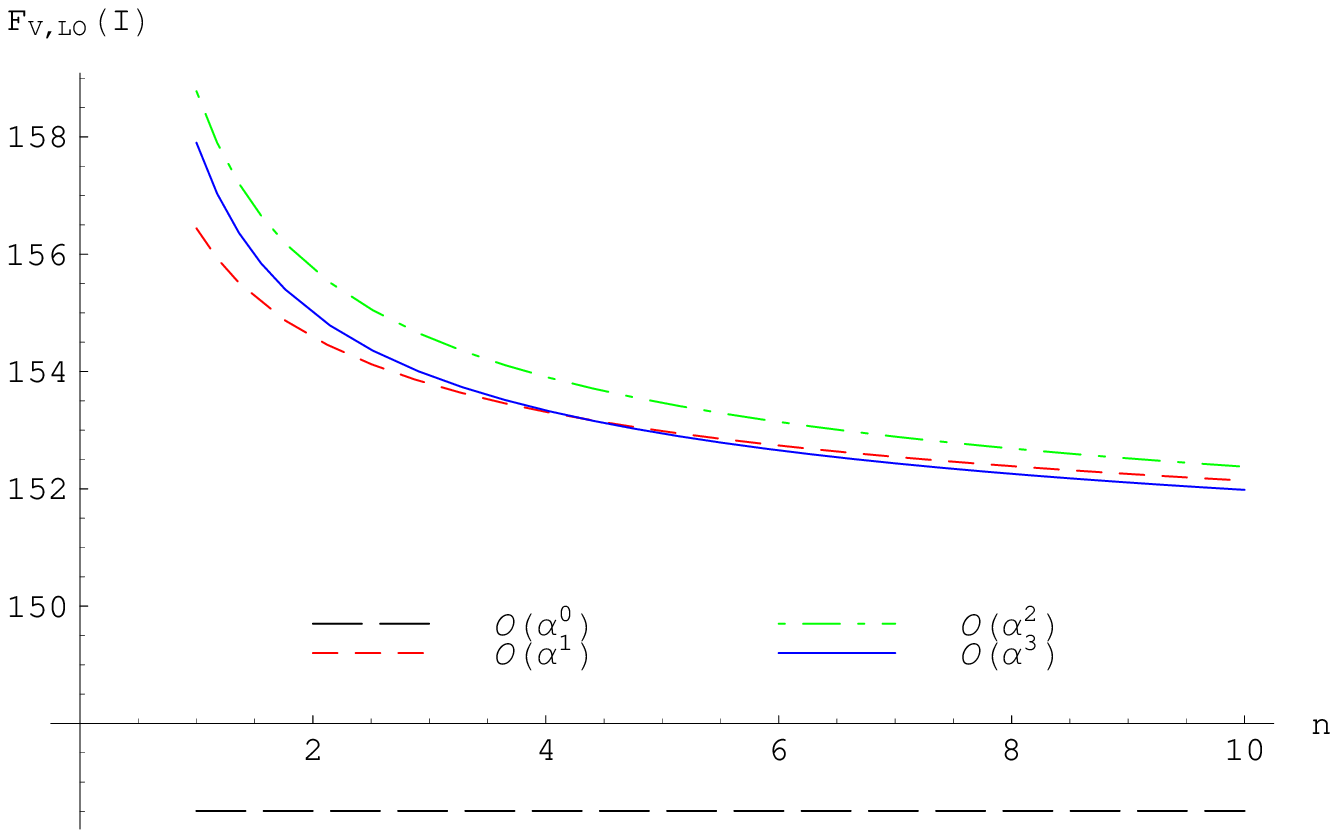}
\hspace{1in}
~\includegraphics[width=0.41\columnwidth]{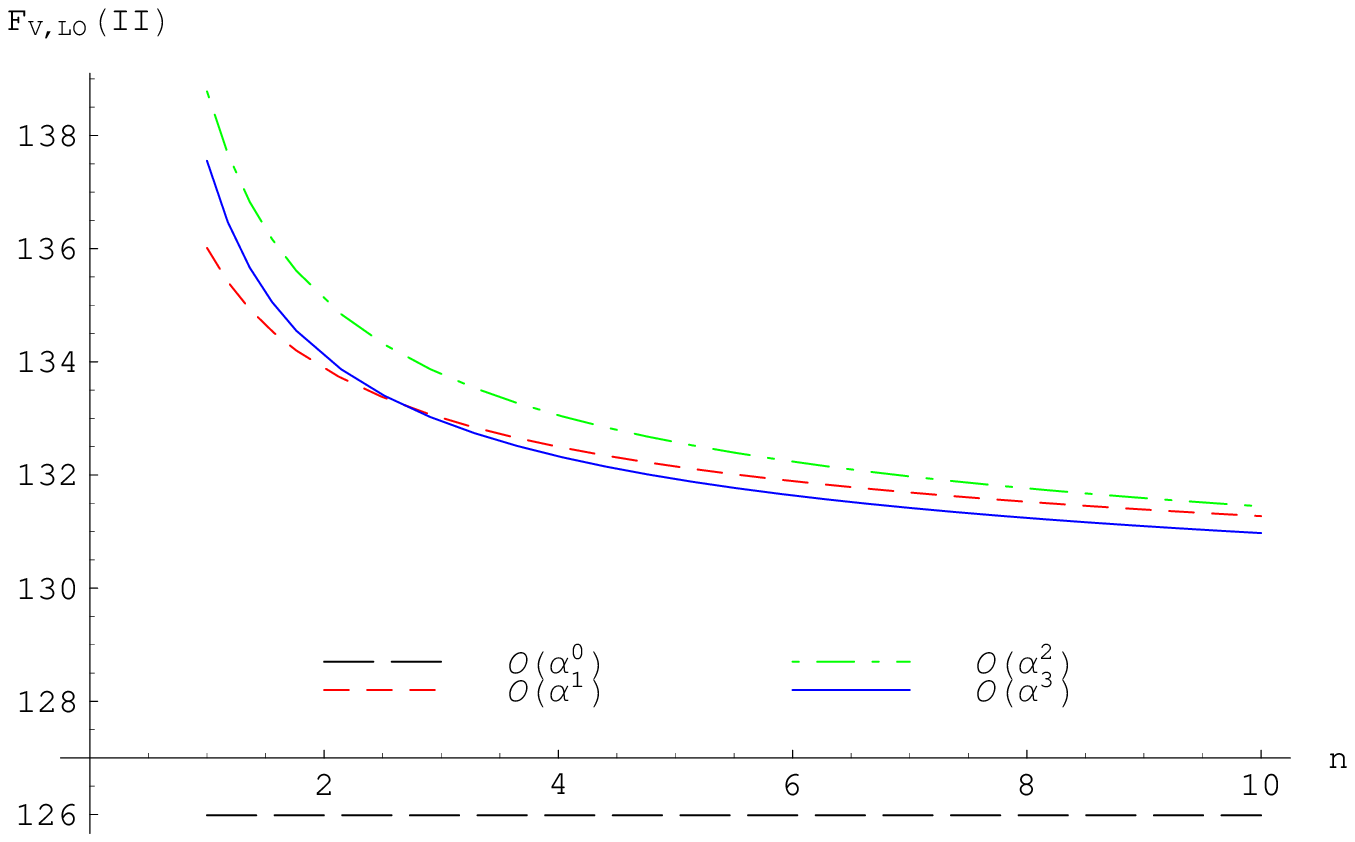}
\caption{\it In this plot we show $F_{V,LO}${\rm (I)} and $F_{V,LO}${\rm (II)} at different orders in $\als$.}
\label{alpha}
\end{figure}

We should mention that the values 
obtained for these parameters are not very stable under the change of number of data points, 
except for $B_V$ and $B_A$, which are roughly stable, although with quite sizeable uncertanties. For the 
subleading terms $A$ and $C$, their values are basically random with the fit. We roughly find $B_V \simeq B_A$ 
within the uncertainties. The $n$ dependence of the axial and vector (model II) decay constants is small but sizeable (and it goes 
in the right direction for low $n$). The $1/n$ corrections 
are always corrections compared with the leading order terms. Nevertheless, the $1/n^2$ 
correction is much larger than the 
$1/n$ one for the range of values of $n$ that we explore. This appears to be due 
to the $\als^2/(4\pi)^2$ suppression of the $1/n$ term, as well as to the difference in size between the constants 
$A$ and $C$. This is so for the axial and vector (model II) decay constants. Nevertheless, for the vector (model I) 
decay constants the $n$ dependence appears to be quite small also at NNLO. This appears to be due to the small value of the 
coefficient $C_V$(I). The gluon condensate contribution is a small correction to the total NNLO term. 
Either way, our predictions compare favorably with experiment when this comparison is possible.

We should keep in mind that these results have been obtained for a specific model, so 
we are testing the impact of the perturbative corrections for this specific model. On the other hand, if 
one believes that the large $n$ behavior of the spectrum is dictated by the Regge behavior and that 
the corrections can be obtained as an expansion in $1/n$, the set up is general. The only ambiguity 
comes from where the logarithms should be introduced (masses or decays). At this respect it is worth mentioning 
that, as a matter of principle, this ambiguity could be fixed if enough experimental information 
were available for the masses and decays.

\section{Conclusions}

We have studied the constraints that the OPE 
imposes on large $N_c$ inspired QCD models for current-current 
correlators. We have focused on the constraints obtained by going 
beyond the leading-order parton computation. 
We have explicitly showed that, assumed a given mass spectrum (Regge plus corrections 
in $1/n$), we can obtain the logarithmic (and constant) behavior in $n$ of the decay constants within a systematic expansion in $1/n$.
More than that, power-like $1/n$ corrections can only be incorporated in the analysis if full consideration 
to the perturbative corrections in the Euclidean regime is made. This is due to the fact that these type of contributions 
produce logarithms of $Q$ in the Euclidean (this is one of the reasons why this sort of corrections are not usually considered in 
quark-hadron duality analysis). On the other hand,  the existence of $\ln n$ in the decay constants may point to the existence of two scales in the problem,
$\lQ$ and $n\lQ$, in the Minkowski regime.
 
We have also performed some numerical estimates of the importance of these corrections. 
The $n$ dependence of the decay constants is small but sizeable for the axial and vector (model II) 
channel, for the vector (model I) one this dependence is small. On the 
other hand the uncertainties of the calculation are large.  Either way, our predictions 
compare favorably with experiment when this comparison is possible.

Our example shows that it is possible to have different large $n$ behavior for the 
vector and pseudo-vector mass spectrum and yet comply with all the constraints from the
OPE. 
  
An important caveat of our analysis is that 
we have not considered what the effect of renormalons could be. We have focused on the effect 
of low orders in perturbation theory to the decay constants. It would be interesting to see 
whether the knowledge of the higher order behavior of perturbation theory may give some 
extra constraints on the values of these constants and the mass spectrum. At this respect we have to say 
that we have obtained approximated expressions for the decay constants as an expansion 
in $\als(nB_V)$, with just the low order contributions in $\als$. It is quite likely that this expansion 
is asymptotic and that different orders in $1/n$ are related in a similar way to the one 
found in the renormalon analysis for the OPE expansion for different orders in $1/Q^2$.
Therefore, the results obtained for the $1/n$ corrections could be affected as well by 
the asymptotic behavior of the $1/\ln n$ expansion in the leading-order term. 
This is obviously related with renormalons. We expect to come back to this issue in the future. 

\medskip

{\bf Acknowledgments}. We thank S. Afonin, A. Andrianov, D. Espriu, and S. Peris for discussions and 
L. Glozman for correspondence. This work is partially supported by the
network Flavianet MRTN-CT-2006-035482, by the spanish grant FPA2004-04582-C02-01, by the catalan grant SGR2005-00564 and by a
{\it Distinci\' o} from the {\it Generalitat de Catalunya}. 


\end{document}